\documentclass[%
 reprint,
 amsmath,amssymb,
 aps,
]{revtex4-2}

\usepackage{graphicx}
\usepackage{dcolumn}
\usepackage{bm}

\usepackage{xcolor}

\begin{document}

\preprint{APS/123-QED}

\title{Divergence of the variance of the optical phase in gain-switched semiconductor lasers described by stochastic rate equations}

\author{Angel Valle}
\affiliation{%
Instituto de F\'\i sica de Cantabria (CSIC-Univ. Cantabria), Avda. Los Castros s/n, E39005, Santander, Spain \\
 }
\date{\today}

\begin{abstract}
In this paper, we report a theoretical study of the phase diffusion  in a gain-switched single-mode semiconductor laser. We use stochastic rate equations for the electrical field to analyze the phase statistics of the gain-switched laser. Their use avoid the instabilities obtained with rate equations for photon number and optical phase when the photon number is small. However we show that a new problem appears when integrating with the field equations: the variance of the optical phase becomes divergent. This divergence can not be observed with the numerical integration of the commonly used equations for photon number and optical phase because of the previous instabilities. The divergence of the phase variance means that this quantity does not reach a fixed value as the integration time step is decreased. We obtain that the phase variance increases as the integration time step decreases with no sign of saturation behaviour even for tiny steps. We explain the divergence by making the analogy of our problem with the 2-dimensional Brownian motion. The fact that the divergence appears is not surprising because already in 1940 Paul L\`evy demonstrated that the variance of the polar angle in a 2-dimensional Brownian motion is a divergent quantity. Our results show that stochastic rate equations for photon number and phase are not appropriated for describing the phase statistics when the photon number is small. Simulation of the stochastic rate equations for the electrical field are consistent with L\`evy's results but gives unphysical results since an infinite value is obtained for a quantity that can be measured.

\end{abstract}

\maketitle


\section{\label{sec:level1}Introduction}

Semiconductor lasers are normally used as single
photon sources in most commercial and research quantum key distribution (QKD) systems \cite{xu2020secure}. Light pulses with random phases are generated by gain-switching these lasers because of the random character of the phase of the spontaneous emission photons that seed these pulses during their formation. Weak coherent pulses (WCP), obtained from attenuation of semiconductor laser pulses, are used as single photon sources in practical QKD systems from early 1990s \cite{paraiso2021advanced}. Another practical application of random-phase pulses emitted by gain-switched semiconductor lasers is quantum random number generation (QRNG) \cite{stipvcevic2014true,ma2016quantum,
herrero2017quantum,mannalath2022comprehensive}. QRNGs are a particular case of hardware physical random number generators in which the data are obtained from quantum events. Their main advantage is that the generated randomness is inherent to quantum mechanics making quantum systems a perfect source of entropy for random number generation \cite{herrero2017quantum}. Applications of QRNGs include those typical of random number generators like Monte Carlo simulations, weather prediction, industrial testing, gambling, quantitative finance, etc. Specific applications of QRNGs can be found in fundamental physics tests and particularly in quantum communications because using these generators is a necessary security requirement for QKD \cite{paraiso2021advanced}.

Most of the existing QRNGs are based on quantum optics because of the availability of high-quality optical components and the possibility of chip-size integration \cite{ma2016quantum}. Single-photon 
\cite{stefanov2000optical,jennewein2000fast,wei2009bias}
 and multiphoton QRNGs 
 \cite{guo2010truly,shen2010practical,qi2010high,
 jofre2011true,argyris2012sub,xu2012ultrafast,
 abellan2014ultra,yuan2014robust,marangon2018long,septriani2020parametric,
 shakhovoy2020quantum,shakhovoy2021influence, abellan2016quantum,lovic} 
 have been demonstrated. QRNGs based on gain-switching semiconductor lasers are an example of multiphoton QRNGs. They exploit the fact that the product of the interference of two pulses with random phases is a third pulse with random amplitude 
 \cite{paraiso2021advanced,
 abellan2014ultra,yuan2014robust,marangon2018long,septriani2020parametric,
 shakhovoy2020quantum,shakhovoy2021influence, abellan2016quantum,lovic}. Advantages of these type of QRNGs include fast operation at Gbps rates, robustness, low cost, operation with flexible clock frequencies, use of standard photodetectors due to the high signal level,  
 and full integration on an InP platform \cite{abellan2016quantum}.
In these generators the current applied to a single-mode laser diode is periodically modulated from a well below threshold value to a value above threshold for obtaining gain-switching operation \cite{jofre2011true,
abellan2014ultra,yuan2014robust,
marangon2018long,
abellan2016quantum,shakhovoy2021influence,lovic}. While the laser is below threshold the evolution of the optical phase becomes random due to the spontaneous emission noise. The laser emits then a series of pulses with a random phase. Phase fluctuations are converted into amplitude fluctuations by using an
unbalanced Mach-Zehnder interferometer, with a delay matching the pulse repetition period. 
Detection and post-processing of the amplitude values provide the generation of random numbers. Fast generation rates, up to 43 Gbps quantum random bit generation, have been experimentally shown \cite{abellan2014ultra}.

The above mentioned generators belong to the class of trusted-device QRNGs \cite{ma2016quantum}. In these systems it is very useful to build a model of the physical entropy source to guarantee unpredictibility, in the sense that the device is generating randomness of genuine quantum origin \cite{lovic}. By using numerical simulations of the stochastic rate equations that quantify the phase noise, a comparison with the experimental results is performed for validating the operational limits of the phase-noise QRNG \cite{lovic}.
This validation process can be used to check the device performance in order to detect malfunctioning or malicious manipulation of the QRNG \cite{lovic}. A good quantitative description of experimental phase noise using stochastic rate equation modelling can only be obtained when extraction of the parameters of the semiconductor laser is performed \cite{quirce2021phase,lovic}.

Experimental and theoretical studies of laser light fluctuations began in late 1960s \cite{lax1966quantum,lax1969quantum,risken1996fokker,henry1996quantum,arecchi1967time}. Study of fluctuations of the light emitted by semiconductor lasers has also received a lot of attention  \cite{coldren2012diode,agrawal2013semiconductor,
spano1983phase,schunk1986noise,henry1986phase,petermann1991laser,
balle1991statistics,
balle1992statistics}. A theoretical description, deduced from first principles and valid for below and above threshold operation, has been based on the Fokker-Planck equation or alternativelly on stochastic rate equations of Langevin's type \cite{risken1996fokker,gardiner1985handbook}. The dynamical description of the statistics of the optical phase in gain-switched semiconductor lasers, that are the basis of phase-noise QRNGs, has been performed using these Langevin's equations
\cite{paraiso2021advanced,lovic,abellan2014ultra,septriani2020parametric,
shakhovoy2021influence,
shak,
quirce2021phase,
quirce2022spontaneous} that correspond to the widely used models described in \cite{coldren2012diode,agrawal2013semiconductor,schunk1986noise,henry1986phase}.   
Quantifying the phase noise in gain-switched lasers is also important in the context of QKD where phase randomization is essential to security  \cite{paraiso2021advanced,lovic,lovic2022quantifying}. 

Most of the previous stochastic rate equations consider the evolution for the carrier and photon densities inside the device and the optical phase of the laser \cite{paraiso2021advanced,lovic,abellan2014ultra,septriani2020parametric,
shakhovoy2021influence,shak}. Numerical integration of these equations using explicit methods like first-order Euler-Maruyama's, second-order Milstein's or implicit methods like Heun's predictor-corrector algorithm present numerical instabilities when the laser is below threshold \cite{quirce2022spontaneous}, that is precisely the regime in which most of the phase randomization occurs. 
When the photon number is very small, random fluctuations that model the spontaneous emission noise can produce a negative value of the photon number. Numerical instabilities appear because the photon number appears inside the square root factors that multiply the noise terms in the equations for photon number and optical phase.
These instabilities can be avoided when using the corresponding rate equations for the laser electrical field and carrier number \cite{quirce2022spontaneous}. 

In this work we will use these E-field equations for analyzing the phase statistics of the gain-switched laser. As mentioned above, their use avoid the previously described instabilities but we will show that a new problem appears: the variance of the optical phase becomes divergent. The divergence can not be observed with the simulation of the equations for photon number and optical phase because the appearance of the previous instabilities does not permit the correct calculation of the dynamical evolution when the power is very small. We find that the phase variance divergence manifests by a monotonous increase of its value as the integration time step decreases. We compare the optical phase evolution obtained with the stochatic rate equations with that obtained in a 2-dimensional Brownian motion. In this way we use the long known result of the divergence of the variance of the polar angle in the plane Brownian motion \cite{levy1940mouvement} to explain the divergence of the phase variance. 
 
Our paper is organized as follows. In section 2, we present our theoretical model. Section 3 is devoted to the present our numerical results with a special emphasis on the divergence of the phase variance. In section 4, we discuss the origin of this divergence. Finally, in section 5 we discuss and summarize our results.

\section{Theoretical model}

The dynamics of a gain-switched single-mode laser diode can be modelled by using a set of stochastic differential rate-equations. These read (in Ito's sense) \cite{coldren2012diode,schunk1986noise,petermann1991laser,
paraiso2021advanced,lovic,abellan2014ultra,septriani2020parametric,
shakhovoy2021influence,
shak,
quirce2021phase,rosado2019numerical,quirce2022spontaneous}

\begin{align}
\frac{d p}{d t} &= \left[ \frac{\Gamma v_g g(n)}{1+\bar{\epsilon} p}  -\frac{1}{\tau_p} \right] p + \Gamma R_{sp}(n)\nonumber
\\ &~~~~ + \sqrt{2 \Gamma R_{sp}(n) p} F_p (t) \label{eq:pm} \\
	\frac{d \phi}{d t} &= \frac{\alpha}{2} \left[ \Gamma v_g  g(n)  -\frac{1}{\tau_p} \right] + \sqrt{\frac{\Gamma R_{sp}(n)}{2p} } F_\phi (t)  \label{eq:phim} \\
   \frac{d n}{d t} &= \frac{I(t)}{e V_{a}} - R(n) - \frac{v_g g(n)p}{1+\bar{\epsilon} p}  \label{eq:nm}	
\end{align}

where 
$p(t)$ is the photon density, $\phi (t)$ is the optical phase in the reference frame corresponding to the resonant frequency at the threshold current \cite{petermann1991laser}, and $n(t)$ is the carrier density. In these equations $V_{a}$ is the active volume, $e$ the
electron charge, $v_g$ the
group velocity, $g(n)$ the material gain, $\bar{\epsilon}$ the non-linear gain
coefficient, $\Gamma$ the optical confinement factor, $\tau_p$ the photon
lifetime, and $\alpha$ is the linewidth enhancement factor. $R(n)$ stands for the carrier recombination rate and 
$R_{sp}(n)$ is the rate of the spontaneous emission coupled into
the lasing mode. We consider a temporal dependence for the injected current, $I(t)$, and a material gain, $g(n)$, given by $g(n) = \frac{dg}{dn} (n-n_{t})$,
where $\frac{dg}{dn}$ is the differential gain and $n_{t}$ the transparency carrier density. The Langevin terms $F_p(t)$ and $F_\phi(t)$ in Eqs.~(\ref{eq:pm})-(\ref{eq:phim}),
represent fluctuations due to spontaneous emission,
with the following correlation properties,
$< F_i(t) F_j(t^\prime)> = \delta_{ij} \delta (t-t^\prime)$,
where $\delta (t)$ is the Dirac delta function and $\delta_{ij}$ the Kronecker delta function
with the subindexes $i$ and $j$ referring to the variables $p$ and $\phi$.  We have not taken into account in our model the carrier noise terms because it has been shown that their effect on the statistics of the phase is very small during transient regimes    
 \cite{balle1991statistics,septriani2020parametric,valle2021statistics}.

We now write the corresponding equations for the number of photons inside the laser, $P(t)$, and the number of carriers in the active region, $N(t)$ by doing the following change of variables: $P=pV_p$, $N=nV_{a}$, where $V_p$ is the volume occupied by the photons. These equations read:

\begin{align}   
	\frac{d P}{d t} &=  \left[ \frac{G_N(N-N_t)}{1+\epsilon P}  -\frac{1}{\tau_p} \right] P + \beta B N^2 \nonumber
	\\&~~~~ + \sqrt{2 \beta BP}N F_p (t) \label{eq:P} \\
	\frac{d \phi}{d t} &= \frac{\alpha}{2} \left[ G_N(N-N_t) -\frac{1}{\tau_p} \right] + \sqrt{\frac{\beta B}{2P}} N F_\phi (t)  \label{eq:phi} \\
	\frac{d N}{d t} &= \frac{I(t)}{e} - (AN+BN^2+CN^3)- \frac{G_N (N-N_t)P}{1+\epsilon P}  \label{eq:N} 
\end{align}

For obtaining  these equations we have considered that $R(n)=an + bn^2 + cn^3$ and $R_{sp}(n)=\beta bn^2$, where 
$a, b$ and $c$ are the non-radiative, spontaneous, and
Auger recombination coefficients, respectively, and $\beta$ is the fraction of spontaneous emission coupled into the lasing mode. The expressions of the new parameters are $G_N=\Gamma v_g\frac{dg}{dn}/V_{a}$, $N_t=n_tV_a$, $A=a$, $B=b/V_{a}$, $C=c/V_{a}^2$, $\epsilon =\bar{\epsilon}\Gamma /V_{a}$. In deriving these equations we have also used that $\Gamma = V_{a}/V_p$.

When laser diodes are used for phase-noise QRNGs a large signal modulation of $I(t)$ is considered in such a way that a random evolution of the phase is induced by the spontaneous emission noise, specially when the bias current is below the threshold current, $I_{th}$. Information about the temporal dependence of the phase statistics under large signal current modulation has been obtained by numerical solution of  
Eqs.~(\ref{eq:pm})-(\ref{eq:nm}) \cite{lovic,shak} by using the Euler-Maruyama method \cite{risken1996fokker,kloeden1992stochastic}. As discussed in the previous section numerical integration of these equations is problematic:  when $P$ is very small negative values of $P$ can appear in the square root factors that multiply the noise terms in Eqs.~(\ref{eq:P})-(\ref{eq:phi}) causing instabilities. The usual solution for this problem has been the integration of the corresponding rate equations for the complex electric field, $E$, instead of equations for $p$ and $\phi$ \cite{balle1993parametric,quirce2022spontaneous}. 
These equations read \cite{quirce2022spontaneous}
 
\begin{align}
	\frac{d E}{d t} &= \left[ \left(\frac{1}{1+\epsilon \mid E\mid^2}+i\alpha\right)G_N(N-N_t)  -\frac{1+i\alpha}{\tau_p} \right] \frac{E}{2} 
	\nonumber
	\\&~~~~ +  \sqrt{\frac{\beta B}{2}}N \xi (t) \label{eq:Efield}\\	
	   \frac{d N}{d t} &= \frac{I(t)}{e} -(AN+BN^2+CN^3) - \frac{G_N (N-N_t)\mid E\mid^2}{1+\epsilon \mid E\mid^2}  \label{eq:Nfield} 
\end{align} 
 
where $E(t)=E_1(t)+iE_2(t)$ is the complex electric field and $\xi (t)=\xi_1 (t)+i\xi_2(t)$ is the complex Gaussian white noise with zero average and correlation given by $< \xi(t) \xi^*(t^\prime)> = 2\delta (t-t^\prime)$ that represents the spontaneous emission noise. The  application of the rules for the change of variables in the Ito's calculus \cite{gardiner1985handbook} to $P=\mid E\mid ^2=E_1^2+E_2^2$ and $\phi = \arctan {(E_2/E_1)}$ in Eqs.~(\ref{eq:Efield})-(\ref{eq:Nfield}) gives our initial Eqs.~(\ref{eq:P})-(\ref{eq:N}), as it is explained in Appendix A. Integration of Eqs.~ (\ref{eq:Efield})-(\ref{eq:Nfield}) avoids the previously mentioned instabilities because $P$ does not appear inside the square root factors that multiply the noise terms and hence no instabilities are observed. We include in Appendix B the equations corresponding to the implementation of the Euler-Maruyama method to our model. We also discuss in that appendix the numerical procedure used to obtain the optical phase from the integration of the equations for the real and imaginary part of the electrical field.

\section{Numerical results}

In this section we numerically solve Eq.~(\ref{eq:Efield}) and Eq.~(\ref{eq:Nfield}) by using the Euler-Maruyama algorithm \cite{risken1996fokker,kloeden1992stochastic}. We use the numerical values of the 
parameters that have been extracted for a discrete mode edge-emitting   laser \cite{rosado2019numerical,quirce2022spontaneous}. This device is a single longitudinal mode semiconductor laser emitting close to 1550 nm wavelength and $I_{th}=14.14$ mA at a temperature of 25$^o$C. The values of  the parameters are $G_N=1.48\times 10^4 $s$^{-1}$, $N_t=1.93\times 10^7$, $\epsilon = 7.73\times 10^{-8}$, $\tau_p = 2.17 $ ps, $\alpha =3$, $\beta =5.3\times 10^{-6}$, $A=2.8\times 10^8$ s$^{-1}$, $B=9.8$ s$^{-1}$, and $C=3.84\times 10^{-7}$ s$^{-1}$ \cite{rosado2019numerical,quirce2022spontaneous}. Simulation and experimental results have shown not only qualitative but also a remarkable quantitative agreement for a very wide range of gain-switching conditions \cite{rosado2019numerical,quirce2020nonlinear,quirce2021phase,
quirce2022spontaneous}.

We consider an injected bias current that follows a square-wave modulation of period $T$ with $I(t)=I_{on}$ during $T/2$, and $I(t)=I_{off}$ during the rest of the period. We take the following values: $I_{on}=30 $ mA, and $T=1$ ns. The laser is switched-off with $I_{off}$ below the threshold value in order to get randomness of the phase due to the spontaneous emission noise. The temporal evolution during several consecutive periods of $P$, $N$, and $\phi$ is plotted in Fig. 1 of \cite{valle2021statistics} when $I_{off}=7$ mA. Similar evolutions but plotted in one time window of duration $T$ can be found in Fig. 5 of \cite{quirce2022spontaneous}. In this figure the initial conditions for one period correspond to the final values of the variables at the end of the previous period. The final value of the phase at the end of the period, $\phi (T)$, leaves the $[0,2\pi )$ interval, so the initial value of the phase, $\phi (0)$, must converted to the $[0,2\pi )$ range if we want to obtain well defined statistical moments of $\phi$ \cite{quirce2022spontaneous,valle2021statistics}. This is done by taking $\phi (0)$ as $\phi (0)=\phi (T)-$ int$\bigl(\frac{\phi (T)}{2\pi}\bigr)2\pi$ \cite{quirce2022spontaneous,valle2021statistics}. 

Figs. 1(a), 1(b), and 1(c) show the dynamical evolution of the averaged photon number, carrier number, and optical phase, respectively. The rate equations parameters and the integration time step (0.01 ps) are similar to the values used in \cite{quirce2022spontaneous}. The effect of spontaneous emission noise on some individual realizations under the same modulation conditions of Fig. 1 is well illustrated in Fig. 5 of \cite{quirce2022spontaneous} with its corresponding discussion. Fluctuations of $P$ and $\phi$ are more important at the beginning and at the end of the period (before 0.1 ns and after 0.7 ns, respectively). Since $P$ is small in those regions the noise terms dominate in Eq.~(4) and Eq.~(5) \cite{quirce2022spontaneous}. Fig. 1(d) shows the dynamical evolution of the standard deviation of the phase, $\sigma_\phi$,  that is the relevant quantity for determining the performance of the gain-switched laser diode as a QRNG. The initial value is larger than zero due to our choice of random initial conditions. The two large increases of $\sigma_\phi (t)$ occur at the beginning and at the end of the period and correspond to a phase diffusion regime because in those regions $P$ has small values determined by the spontaneous emission noise.  

\begin{figure}[h!]
\centering
\includegraphics[trim=0 50 0 40, clip,width=9cm]{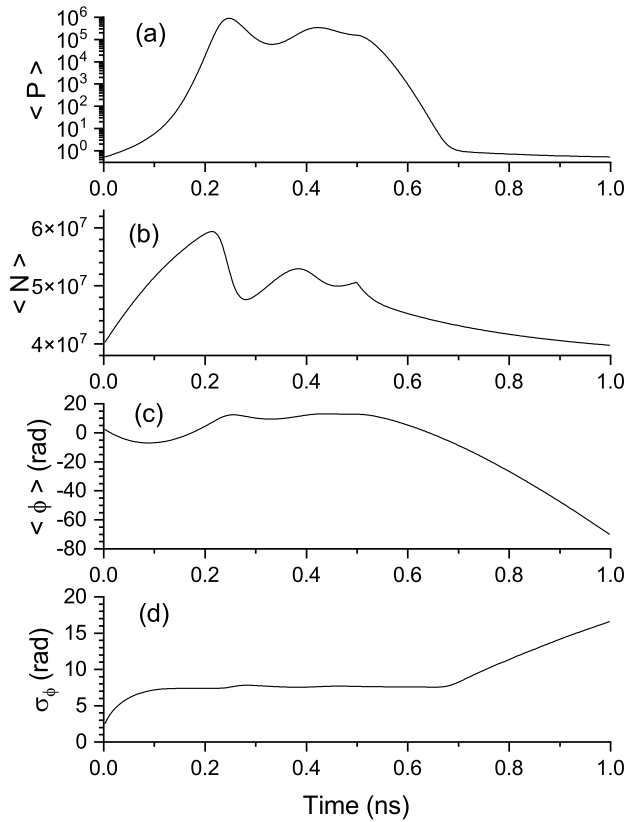}
\caption{(a) Averaged photon number, (b) averaged carrier number, (c) averaged phase, and (d) standard deviation of the phase as a function of time. In this figure $I_{off}=7$ mA, the number of periods is $5\times 10^4$ and the integration time step is 0.01 ps.}
\end{figure}

Fig. 2 shows the effect of decreasing the integration time step , $\Delta t$, on the dynamical evolution of the phase statistics. A change of several orders of magnitude, from $\Delta t=10^{-1}$ ps to $\Delta t=10^{-6}$ ps is considered. Fig. 2(a) shows that the averaged phase quickly achieves convergent values because it does not change significatively with $\Delta t$. A similar convergence (not shown) occurs with the averaged values and standard deviations of $P$ and $N$. However, the behaviour of the variance of the optical phase, $\sigma_\phi^2 (t)$, is radically different, as it is shown in Fig.~2(b). There is not any evidence of the convergence of this quantity as the time step is decreased, even reaching values of $\Delta t$ as tiny as $10^{-5}$ or $10^{-6}$ ps. The differences between results for different time steps appear at the beginning and at the end of the period. In these regions linear increases of $\sigma_\phi^2 (t)$ with $t$ appear with slopes that increase as $\Delta t$ is decreased. The linear increase of the phase variance is characteristic of the phase diffusion process. The fact that the slope of $\sigma_\phi^2 (t)$ during the phase diffusion regime keeps on increasing when $\Delta t$ decreases indicates that the phase variance is a divergent quantity: its value can be arbitrary large providing that a small enough integration time step is considered.

\begin{figure}[h!]
\centering
\includegraphics[trim=0 50 0 40, clip,width=9cm]{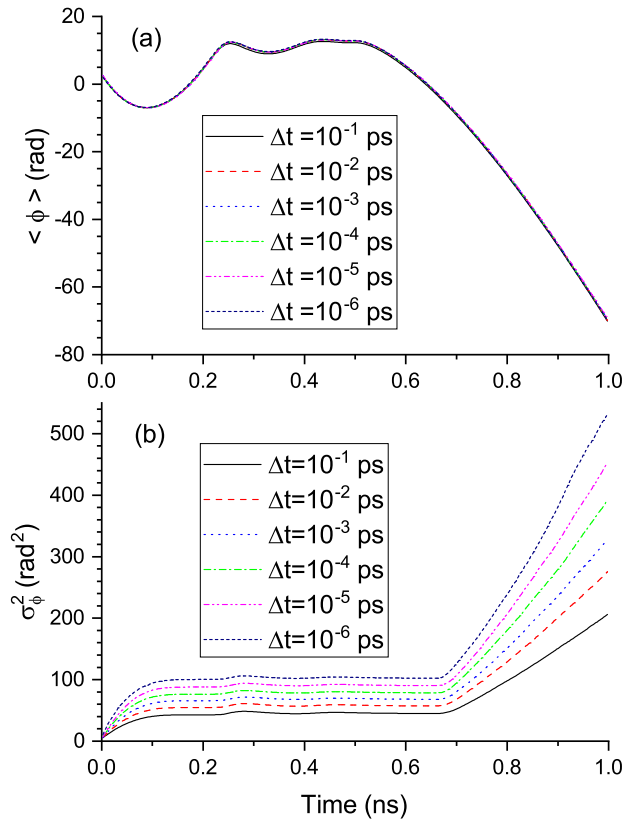}
\caption{(a) Averaged phase, and (b) variance of the phase as a function of time. Results for different integration time steps are plotted with lines of different colours, as indicated in the legend. In this figure $I_{off}=7$ mA and the number of periods is $5\times 10^4$.}
\end{figure}

To show that there is not any sign of convergence in the values of the previous slopes, and in consequence of the phase variance values, we have calculated the slope of $\sigma_\phi^2$ versus $t$ at the beginning and at the end of the period.
These results are shown in Fig. 3 for the different $\Delta t$ considered in Fig.~2. The initial (final) slope has been calculated by linear fitting the values included in Fig.~2 from 0 to 0.025 ns (0.9 to 1 ns). Adjusted $R-$square merit values, $\bar{R}^2$, that give information about goodness of fit, range from 0.99850 to 0.99992, indicating that good linear fits have been obtained. 

\begin{figure}[h!]
\centering
\includegraphics[trim=0 30 0 50, clip,width=9cm]{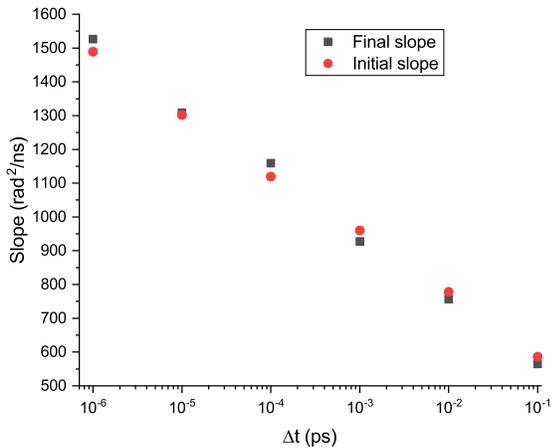}
\caption{(a) Slope of the phase variance vs. time as a function of the integration time step. Results for the final and initial part of the period are plotted with different symbols.}
\end{figure}

The fact that phase diffusion is the dominant process at the beginning and at the end of the period should result in similar slopes in those time regions. Fig. 3 shows that the values of the initial and final slopes are similar. More importantly, 
Fig. 3 also shows that both slopes increase as the time step decreases with no sign of saturation at the smallest values of $\Delta t$. We have plotted Fig.~3 with a logarithmic horizontal axis. Fig.~3 also shows that the slope depends linearly on $\log \Delta t$ so the phase divergence is characterized by a logarithmic dependence of the slope on $\Delta t$: slope=$-c_1\log \Delta t+c_2$, where $c_1, c_2 > 0$, and log is the natural logarithm.
In fact, linear fitting of the data contained in Fig. 3 gives slope=$-77.525\log\Delta t+414.277$, $\bar{R}^2=0.9993$ for the initial slope, and  slope=$-83.062\log\Delta t+371.362$, $\bar{R}^2=0.9976$, for the final slope. The divergence of the phase variance is then characterized by the previous logarithmic dependence because if we calculate the change of $\sigma_\phi^2$ during a time interval $(t_1,t_2)$ in which  phase diffusion dominates the evolution we obtain that $\sigma_\phi^2 (t_2)-\sigma_\phi^2 (t_1)=(-c_1\log \Delta t+c_2)(t_2-t_1)$.

\section{Explaining the divergence}

We will explain the divergence of the phase by using the analogy of our problem, under certain restrictions, to the simplest 2-dimensional Brownian motion, that corresponds to a particle diffusing in a plane subject only to the random force. Let us first consider a semiconductor laser biased slightly below threshold, $I_{off}\lesssim I_{th}$, and with $\alpha =0$. In this case $E/2$ and the term that multiplies it in Eq.~(\ref{eq:Efield}) are small and the noise term dominates the evolution at the beginning and at the end of the period. In this way we can approximate Eq.~(\ref{eq:Efield}) by:

\begin{equation}
\frac{d E}{d t}=\sqrt{\frac{\beta B}{2}}N \xi(t)\label{eq:Efieldapp}
\end{equation}

Since the real and imaginary part of $\xi(t)$, $\xi_1(t)$ and $\xi_2(t)$, are independent Gaussian white noises, the evolution of the real and imaginary parts of the electric field, $E_1(t)$ and $E_2(t)$, is described with a 2-dimensional Brownian motion. $E_1(t)$ and $E_2(t)$ are independent Gaussian processes, each of them with zero mean value and diffusing with a diffusion coefficient given by $\beta BN^2/2$. We also note that considering $\alpha =0$ means that the phase evolution is only affected by the spontaneous emission noise term as it can be seen in Eq.~(\ref{eq:phi}).

A first clue of the divergent behavior of the phase is given by the distribution of the ratio of those two Gaussians, $E_2/E_1$, since its calculation is an intermediate step to calculate the phase. It is shown in \cite{pham2006density} that the probability density of this ratio is that corresponding to a Cauchy distribution, that is a well known example of continuous random variable with infinite variance. However, the quantity of interest is the phase, $\phi =\arctan(E_2/E_1)$, so we will extend our discussion using the well established mathematical theory of 2-dimensional Brownian motion. 

A freely diffusing particle in two dimensions, that is, one executing  2-dimensional Brownian motion can be described mathematically in polar coordinates. Since $E_1(t)$ and $E_2(t)$ are continuous stochastic processes the polar angle, $\phi (t)$, of the corresponding plane Brownian motion has the freedom to vary from $-\infty$ to $\infty$. $\phi (t)$ is the winding number of the continuous path $E(\tau )=E_1(\tau )+iE_2(\tau )$, $0 \leq \tau \leq t$, about the origin \cite{spitzer}. $\phi (t)$ is a continuous function of $t$ with probability one (the probability is zero that $E(t)$=0 in any $t-$interval) \cite{levy1940mouvement,spitzer}. Since $\phi$ takes values in $(-\infty , \infty )$ instead of $[0, 2\pi )$, we can view the plane Brownian motion as taking place in the universal cover of $\mathbb{C}\setminus\{ 0\}$, that is the Riemann surface of log $z$.
There is a long known but particularly striking feature of polar angle evolution in the plane Brownian motion: $\phi (t)$ has infinite variance, $\sigma_\phi^2 (t) = \infty$, as Paul L\` evy demonstrated \cite{levy1940mouvement} from the fact that $\phi (t)$ tends to assume very large values due to the roughness of the Brownian trajectory when $P(t)$ is close to zero. This arises from the fractal nature, the infinitesimal, infinitely frequent, random walk steps of the plane Brownian motion, and applies whatever the starting radius, $\sqrt{P(0)}$, is from the origin \cite{hannay2022mean}. 

The reason why we are observing the divergence described in the previous section is the infinite value of the variance of the phase. We show in Fig.~4 the evolution of $P(t)$ and $\phi (t)$ for different random trajectories when $\alpha =0$, and $I_{off}=0.9 I_{th}$ (12.73 mA). Results obtained with $\Delta t=10^{-1}$ ps and $\Delta t=10^{-5}$ ps are shown in the left and right parts of the figure, respectively. In each figure we show five typical trajectories and a non-typical trajectory (solid black line) for which the maximum value of $\mid\phi (t)\mid$ was obtained in a simulation with 5$\times 10^4$ periods. It is clear in Fig.~4(a) and Fig.~4(b) how the spontaneous emission noise dominates the evolution at the beginning and at the end of the period. Fig.~4(c) and Fig.~4(d) shows that the realizations of $\phi$ are continuous functions of $t$ in such a way that they are also dominated by noise at the previously mentioned regions. It is clear from  Fig.~4(a) and Fig.~4(c) that the largest excursions of the phase appear in the trajectories that get closer to zero (see the evolution of the solid black lines close to 0.82, 0.85, and 0.91 ns). In these cases there are more rotations induced by the noise around the origin of the complex plane $(E_1,E_2)$.
Fig.~4(b) and Fig.~4(d) show the results obtained when $\Delta t$ has decreased 4 orders of magnitude. The decrease of $\Delta t$ results in more frequent rotations induced by noise around the origin for a given trajectory. This can be seen by comparing the two trajectories with the largest phase excursions of Fig.~4(c) and Fig.~4(d): these excursions widen when $\Delta t$ decreases. In fact the maximum value of the phase  observed for our simulations with $5\times 10^4$ trajectories is $\phi /(2\pi )$= 6.63, 10.44, 11.22, 12.03, and 12.97, for $\Delta t = 10^{-1}, 10^{-2}, 10^{-3},10^{-4} $, and $10^{-5}$ ps, respectively.

\begin{figure}[h!]
\centering
\includegraphics[trim=47 50 0 40, clip,width=10cm]{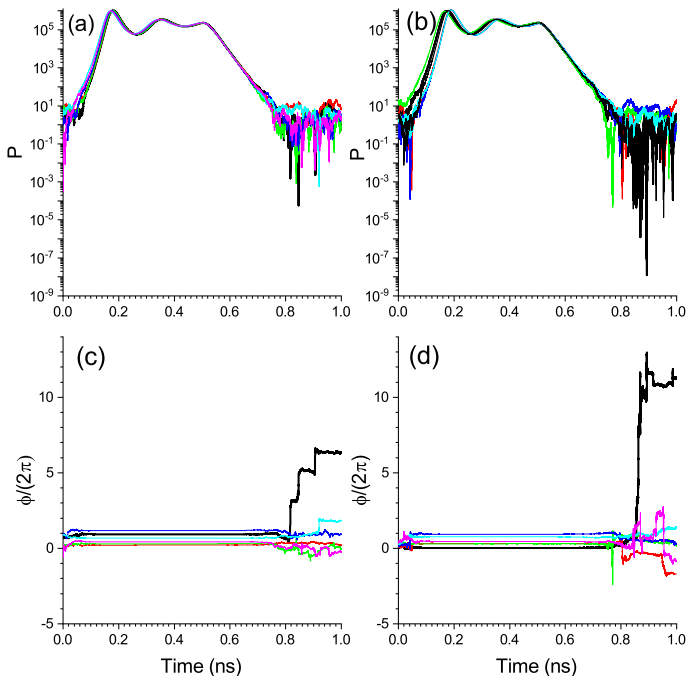}
\caption{Different realizations of the (a)-(b) photon number, and (c)-(d) optical phase as a function of time. The integration time step is $10^{-1}$ ps ($10^{-5}$ ps) in (a),(c) ((b),(d)). In this figure $\alpha=0$, and $I_{off}=12.73$ mA. The trajectory for which the maximum value of $\mid\phi (t)\mid$ was obtained in a simulation with 5$\times 10^4$ periods is plotted with solid black line.}
\end{figure}

Widening of the random trajectories as $\Delta t$ decreases has been previously discussed with those in which the excursions are maxima but also occurs for the other trajectories. Since the averaged phase and variance are calculated using a fixed number of trajectories the previous widening results in an increase of the variance of the phase in the diffusing regions as $\Delta t$ decreases as it can be seen in Fig.~5.   These increases are similar to those shown in Fig.~2(b). The discussion using the visualization of random trajectories has been done for the case of $\alpha =0$ and $I_{off}=0.9 I_{th}$ because the effects of drift in the phase are minimized in such a way that we have a better correspondance with the 2-dimensional free Brownian motion.

\begin{figure}[h!]
\centering
\includegraphics[trim=0 30 0 50, clip,width=9cm]{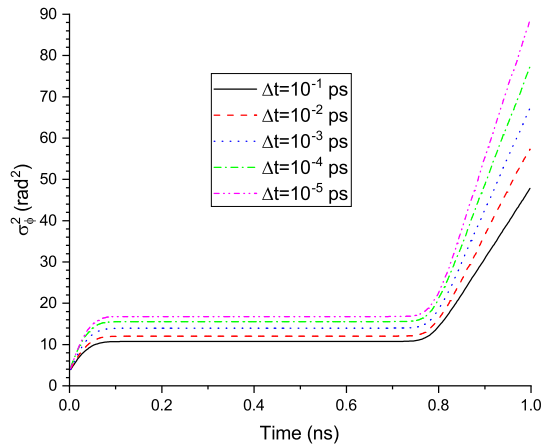}
\caption{Variance of the phase as a function of time. Results for different integration time steps are plotted with lines of different colours, as indicated in the legend. In this figure $\alpha = 0$, $I_{off}=12.73$ mA, and the number of periods is $5\times 10^4$.}
\end{figure}

We now discuss the situation found when having a realistic value of $\alpha$ and a smaller value of $I_{off}$ in order to have a stronger effect of the phase diffusion ($\alpha =3$, $I_{off}= 7$ mA), that is the precisely the case described in the previous section. 
Fig.~6 shows  $P(t)$ and $\phi (t)$ for six different random trajectories obtained with two different values of $\Delta t$ ($10^{-1}$, and $10^{-5}$ ps, at the left and right parts of the figure, respectively).

\begin{figure}[h!]
\centering
\includegraphics[trim=47 65 0 45, clip,width=9cm]{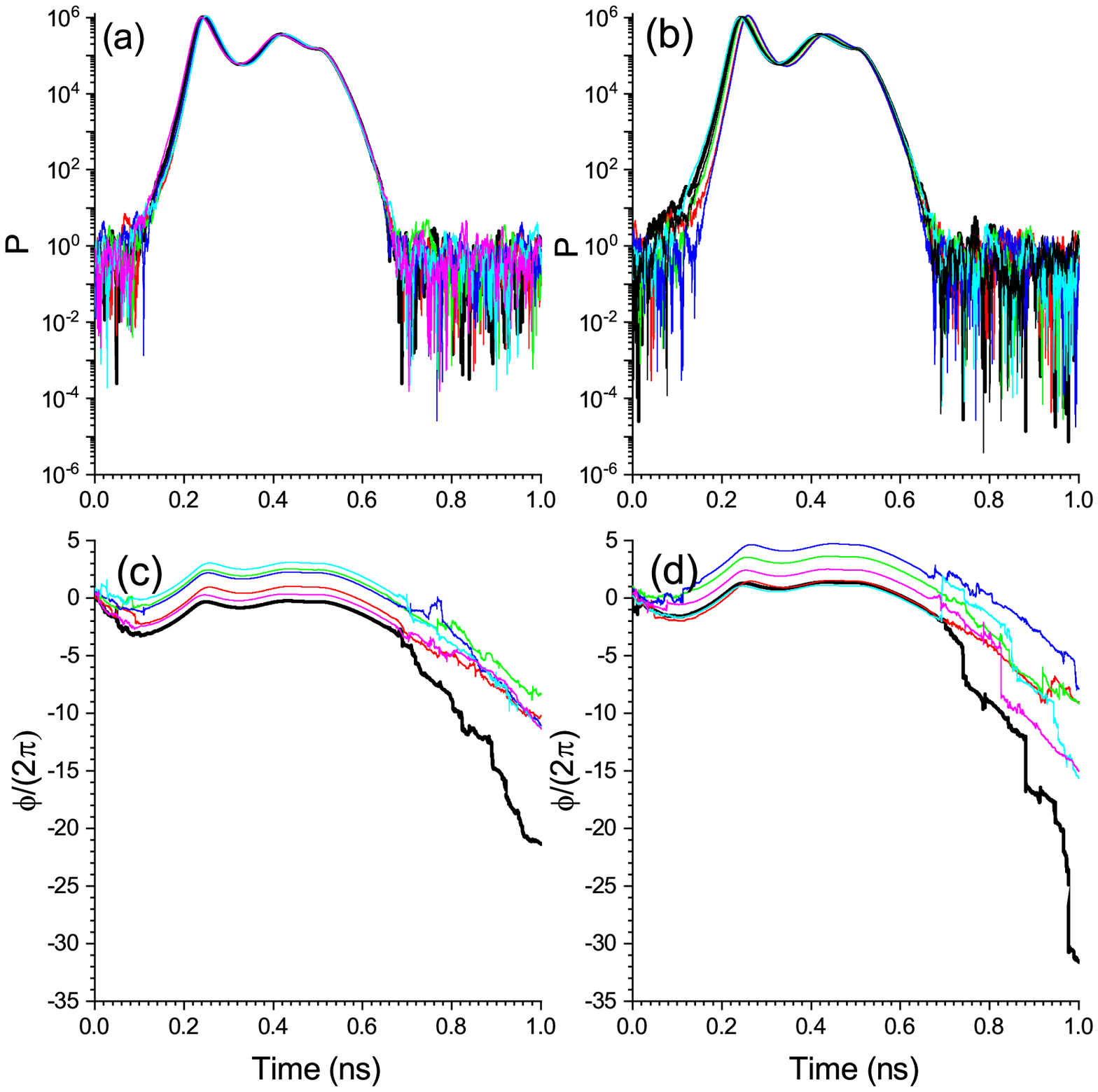}
\caption{Different realizations of the (a)-(b) photon number, and (c)-(d) optical phase as a function of time. The integration time step is $10^{-1}$ ps ($10^{-5}$ ps) in (a),(c) ((b),(d)). In this figure $\alpha=3$, and $I_{off}=7$ mA. The trajectory for which the maximum value of $\mid\phi (t)\mid$ was obtained in a simulation with 5$\times 10^4$ periods is plotted with solid black line.}
\end{figure}

Again we show the trajectory for which the maximum value of $\mid\phi (t)\mid$ was obtained in our simulations and the trajectories corresponding to five consecutive periods.
Fig.~6(c) and Fig.~6(d) show that both deterministic drift and fluctuations of the phase are important in determining the value of the phase at the beginning and at the end of the period. While $P$ is large the phase evolution is mainly deterministic, being characterized by the relaxation oscillations before $t=T/2$ and a monotonous decrease after $t=T/2$ since $N(t) < N_{th}$ (see Eq.~(\ref{eq:phi})). Fig.~6 shows that the largest excursions of the phase appear in the trajectories that get closer to zero (see for instance the evolution of the solid black lines close to 0.74, 0.88, and 0.98 ns in Fig.~6(b) and Fig.~6(d)). Also comparison between  
Fig.~6(a) and Fig.~6(b) shows that when $\Delta t$ decreases the minimum values of $P$ are closer to zero with more frequent rotations induced by noise around the origin of the complex plane. This is seen by the widening of the trajectory with the largest phase excursions in Fig.~6(c) and Fig.~6(d) as $\Delta t$ decreases. Widening of typical random trajectories as $\Delta t$ decreases is clearly seen when comparing Fig.~6(c) and Fig.~6(d). Widening of trajectories explains the increase of the variance of the phase in the diffusing regions as $\Delta t$ decreases as it was shown in Fig.~2(b).

\begin{figure}[h!]
\centering
\includegraphics[trim=47 0 0 20, clip,width=9cm]{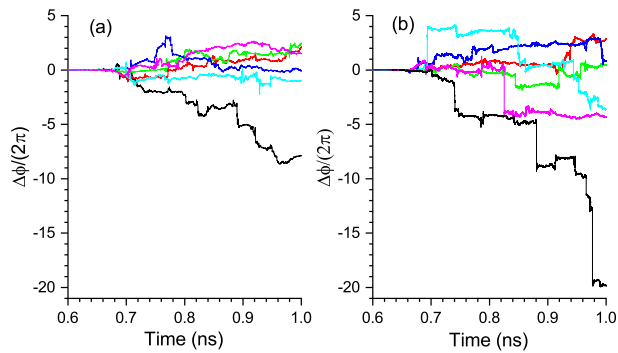}
\caption{Difference between the random phase of the trajectories in Fig. 6 and their corresponding deterministic phase. Each trajectory is identified with the same colour in Fig. 6 and Fig. 7. Fig. 7(a) and Fig. 7(b) correspond to Fig. 6(c) and Fig. 6(d), respectively.
}
\end{figure}

We show in Fig. 7 the difference between the random phase of the trajectories in Fig. 6, $\phi(t)$, and their corresponding deterministic phase, $\phi_{det}(t)$, that is $\Delta\phi (t) = \phi(t)-\phi_{det}(t)$. The values of $\phi_{det}(t)$ have been calculated from the integration of the deterministic rate equations from the initial conditions at $t$=0.2 ns, time at which a deterministic evolution has been reached for all the trajectories. Under these conditions the effect of noise becomes important just before $t$=0.7 ns and the features already discussed in Fig. 6 are better visualized: the large random excursions of the phase at 0.74, 0.88 and 0.98 ns are well seen in Fig. 7(b), and the increasing widening of trajectories as the integration step decreases is also well seen by comparing Fig. 7(a) and Fig. 7(b).

Another way of understanding the divergence of the variance of the phase is by using the following result. Spitzer \cite{spitzer} 
showed that the normalized winding angle obtained with unit diffusivity, $\Theta (t)$, converges to a standard Cauchy distribution as $t\rightarrow\infty$:

\begin{equation}
\frac{2\Theta (t)}{\log t}\xrightarrow{d} X,\,\,\,\,\, f(x)=\frac{1}{\pi}
\frac{1}{1+x^2},
\label{eq:spitzer}
\end{equation}
where $f(x)$ is the probability density of $X$.  This means that given a large $t$, $\Theta (t)$, and so $\phi(t)$ behave as a Cauchy distribution, having therefore an infinite variance. This divergence can be numerically manifested when simulating $X$ by using the method of the cumulative distribution function, $F_X(x)$. For the normalized Cauchy distribution $F_X(x)=1/2+\arctan (x)/\pi$, for $-\infty < x < \infty$. Values of $X$ are numerically obtained by using $x=F_X^{-1} (u)=\tan (\pi (u-1/2))$ where $u$, $0 < u <1$, is distributed according the uniform random variable, $U(0,1)$. As the number of simulated values of $u$ increases, closer values to 0 and 1 are obtained that result in values of $x$ closer to $-\infty$ and $\infty$, respectively. The role of these numbers in determining numerically the average of $X^2$, $<X^2>$, is essential because $<X^2>$ keeps on increasing as the number of simulated $x$ values is increased. 

We complete the discussion on the divergence of the phase variance by including in Appendix C a calculation of this quantity when the system reaches the steady state for a current below the threshold value.  We obtain that $\sigma _\phi ^2$ is proportional to $<\frac{1}{P}>$ that is a divergent quantity because $P$ is an exponential random variable for the below threshold operation.

\section{Discussion and summary}

The numerical integration of Eqs.~(\ref{eq:P})-(\ref{eq:N}) presents numerical instabilities as mentioned earlier. In the simulations of these equations using the Euler-Maruyama scheme the unphysical results due to negative values of $P$ and $N$ have been treated by solving the equations with constraint that $P$ and $N$ are non-negative \cite{shak}. However no details are given in \cite{shak} on how that constraint is implemented. In that work simulations are performed at different integration steps such that $\Delta t < $ 0.1 ps in such a way that the results obtained at different $\Delta t$ were the same for all simulations \cite{shak}. Therefore when using that scheme there is a convergence, but not to the right mathematical solution of Eqs. (4)-(6)  because as we have shown in this work, no convergence is achieved when solving the equivalent field equations that avoid the numerical instabilities. The main reason for this difference is that the integration of Eqs.~(\ref{eq:Efield})-(\ref{eq:Nfield}) describes well the evolution close to $P=0$ given by stochastic rate equation modelling. The fact that there are several mathematical theorems \cite{levy1940mouvement,spitzer} confirming the divergence of the polar angle in 2-dimensional Brownian motion gives us confidence in our results.

A correct numerical integration in the below threshold regime is essential when describing QRNGs and QKDs based on gain-switching laser diodes because that is precisely the regime where phase randomization mainly occurs \cite{lovic, lovic2022quantifying}. As we have shown, integration using stochastic rate equations for the photon number and phase does not describe well the phase statistics below threshold. Simulations must be performed using the equations for the complex field from which we obtain the values of $P$ and $\phi$ at each integration step. However, our results show that the phase variance is a divergent quantity, with a value that increases as the integration step is decreased. This divergence is slow but it is still a divergence, meaning that we can not obtain a well defined value of the phase variance when using stochastic rate equations in the below threshold regime. In this way stochastic rate equations for $P(t)$ and $\phi (t)$ \cite{paraiso2021advanced,lovic,shak,abellan2018quantum,quirce2022spontaneous} or for complex $E(t)$ \cite{quirce2022spontaneous} can not describe well the experimental phase statistics when the photon number is small. 

The main reason of this result lies in the idealized mathematical formulation of the spontaneous emission noise as a white noise, $\xi (t)$, that is a quantity with no correlation at different times and with the rather pathological result that $\xi (t)$ has infinite variance \cite{gardiner1985handbook}. We find that the use of white noise as an idealization of a realistic fluctuating signal causes problems when describing some variables of the system, the optical phase in our case. 

A possible solution of this problem could be modelling the spontaneous emission noise with a realistic version of an almost uncorrelated noise, for instance, an Ornstein-Uhlenbeck process with a very small value of the correlation time. We do not know the order of magnitude of that correlation time but it could be established by comparing experimental measurements of the phase variance, like those performed in \cite{lovic,shak}, with the numerical predictions of the stochastic rate equations for $E(t)$ driven by a coloured noise, instead by a white noise. We would expect a convergent behaviour of the phase variance at integration time steps determined by the correlation time of the noise.

A rigorous treatment of spontaneous emission would require quantization of the electric field. At the most fundamental level the origin of these fluctuations (intensity and phase fluctuations) lies in the quantum nature of the lasing process itself  
 \cite{agrawal2013semiconductor}. A proper description requires a quantum mechanical formulation of the rate equations using quantum Langevin terms \cite{lax1966quantum,gardiner2004quantum} or master equations \cite{gardiner2004quantum}. Stochastic rate equation models (SREM) like those used in this work are approximations that worsen as the photon number decreases. There are two types of SREM. The first one, SREM1, corresponds to that derived from first principles by Lax \cite{lax1969quantum} and Henry \cite{henry1986phase} for a system where the matter and radiation have reached equilibrium. In the derivation a constant bias current is assumed in such a way that the steady-state average values of the variables appear in the terms that multiply $F_p(t)$ and $F_\phi(t)$ in Eqs. (4)-(5) (see Eqs. (19)-(20) in \cite{quirce2021phase} for comparison). Both noises appear as additive noises in SREM1. Using this model, when the steady state corresponding to a bias current below threshold has been reached, the optical phase diffuses with a variance that increases linearly with the time. No divergence of the phase variance is observed because the phase evolution is analogous to a 1-dimensional Brownian motion. 

The situation is different when considering a high-frequency modulation of the bias current. The laser is in a transient regime, the matter-radiation equilibrium has not been achieved yet and the exact form of the spontaneous emission noise terms is unknown \cite{balle1992statistics}. A second type of model, SREM2, is considered in which the steady-state averaged values of the variables in the noise terms are substituted by the corresponding variables in order to analyse transient situations. This approximation is the one considered in this work (corresponding to Eqs. (4)-(5) or to Eq. (7))  and is commonly used in the analysis of phase-noise QRNG 
\cite{paraiso2021advanced,lovic,abellan2014ultra,septriani2020parametric,
shakhovoy2021influence,
shak,
quirce2021phase,
quirce2022spontaneous}. To the best of our knowledge this approximation has not been justified in a rigorous way. The use of this approximation has been successful for describing the experimental results relative to the statistics in transient regimes of quantities related to the laser power, like the turn-on timing jitter. However, our work shows that this approximation fails when analysing the statistics of the optical phase. In this approximation spontaneous emission fluctuations appear as multiplicative noises in SREM2, the phase evolution is analogous to that found in a 2-dimensional Brownian motion and the divergence of the phase variance is observed. We note that these results also hold for a constant bias current when using SREM2, so an analysis of phase fluctuations using SREM1 is more appropriate. This analysis has been performed in \cite{quirce2021phase} in which a good comparison between experimental and theoretical phase variance was obtained while the current is larger than 0.6 times the threshold current. For smaller values of the current the theoretical results clearly underestimate the experimental values so a theoretical modelling better than SREM1 is needed to describe the phase fluctuations.

Another consequence of our results is related to the recently proposed procedure for finding the operational limits of a phase-noise QRNG \cite{lovic}. This method is based on quantifying numerically and experimentally the amount of phase noise randomness produced by a gain-switched laser diode. In the method the intensity distribution at the output of the interferometer of the QRNG is measured. An experimental value of the phase variance is extracted from a fit to the previous intensity distribution. A succesful comparison of this value with the Monte Carlo simulation results is needed to validate the operational limits of the QRNG. Since the results of those simulations depend  on the integration time step the results of that comparison would depend on that step. In order to obtain a better method for comparison we propose a slight modification of the process. This consists on using the full width at half maximum (FWHM) of the phase distribution in the validation method instead of the values of the phase variance. We have shown that the phase converges to a standard Cauchy distribution. This random variable has infinite variance but finite FWHM. Calculation of FWHM would avoid the dependence on the integration time step of the quantities used in the validation method.

Summarizing, we have analyzed theoretically the phase diffusion in a gain-switched single-mode semiconductor laser. By using simulations of the stochastic rate equations for the electrical field we have shown that the variance of the optical phase is a divergent quantity. This result can be explained by using the analogy with the mathematical description of the two-dimensional Brownian motion for which it was shown that the variance of the polar angle is infinite \cite{levy1940mouvement}.
The fact that this divergence is not observed with the simulation of the  photon number and phase equations \cite{shak} indicates that usual simulations of that model are not suitable for describing the phase statistics when the photon number is small: the results of the simulation converge, but not to the right mathematical solution. Simulation of the stochastic rate equations for the electrical field is better because they are consistent with the mathematical results but they still have the problem of giving rise to unphysical results since an infinite value is obtained for the phase variance, quantity that can be experimentally obtained \cite{lovic,shak}. 
Our results have shown that stochastic rate equations models are not appropriated for describing the phase statistics when the photon number is small. A more fundamental theoretical description of spontaneous emission is desirable to better characterize the experimental phase statistics in that regime. Our work has impact on QRNG and QKD based on gain-switching of laser diodes where phase randomization, with its corresponding good theoretical description, are essential to security.

\begin{acknowledgments}
The authors acknowledge financial support from the Ministerio de Ciencia e Innovación, Spain. PID2021-123459OB-C22 MCIN/AEI/FEDER,UE.
\end{acknowledgments}

\appendix
\section{Equivalence between rate equations models}

In this appendix we use the Ito calculus to derive  Eqs.~(\ref{eq:P})-(\ref{eq:phi}) from the rate equation for the complex electric field. We first separate Eq.~(\ref{eq:Efield}) in the equations for the real and imaginary part of $E(t)$, $E_1(t)$ and $E_2(t)$, to obtain:

\begin{equation}
\frac{d E_1}{d t}=A_1+\sqrt{\frac{\beta B}{2}}N \xi_1(t)\label{eq:Efield1}
\end{equation}

\begin{equation}
\frac{d E_2}{d t}=A_2+\sqrt{\frac{\beta B}{2}}N \xi_2(t)\label{eq:Efield2}
\end{equation}

where 

\begin{align}
A_1 &=\left(\frac {G_N(N-N_t)}{1+\epsilon \mid E\mid^2}-\frac{1}{\tau_p}\right)\frac{E_1}{2}
\nonumber
\\& -  \alpha \left( G_N(N-N_t) - \frac{1}{\tau_p}\right)\frac{E_2}{2}
\label{eq:A1}
\end{align}

\begin{align}
A_2 &=\left(\frac {G_N(N-N_t)}{1+\epsilon \mid E\mid^2}-\frac{1}{\tau_p}\right)\frac{E_2}{2}
\nonumber
\\& +  \alpha \left( G_N(N-N_t) - \frac{1}{\tau_p}\right)\frac{E_1}{2}
\label{eq:A2}
\end{align}

and $\xi_1$ and $\xi_2$ are real independent Gaussian noises with $< \xi_i (t)>=0$, and $< \xi _i(t) \xi_j(t^\prime)> = \delta_{ij}\delta (t-t^\prime)$, $i,j=1,2$.

The change of variables is performed by using Ito's formula \cite{gardiner1985handbook}. Given an $n$ dimensional vector $\bf{x}$$(t)$ satisfying the stochastic differential equation

\begin{equation}
d\textbf{x}=\textbf{A}(\textbf{x},t)+\mathbb{B}(\textbf{x},t)d\textbf{W}(t)
\end{equation}

where $d\textbf{W}(t)$ is an $n$ dimensional vector formed by $n$ independent differential Wiener processes, a function of $\bf{x}$, $f(\bf{x})$, satisfies \cite{gardiner1985handbook}

\begin{align}
df(\textbf{x}) &= \sum_iA_i(\textbf{x},t)\partial_if(\textbf{x})dt
\nonumber
\\& +  \frac{1}{2}\sum_{i,j}\bigl[\mathbb{B}(\textbf{x},t)\mathbb{B}^T(\textbf{x},t)\bigr]_{ij}\partial_i\partial_jf(\textbf{x})dt
\nonumber
\\& + \sum_{i,j}\mathbb{B}_{ij}(\textbf{x},t)\partial_if(\textbf{x})dW_j(t).
\label{eq:Ito}
\end{align}

In our 2-dimensional case 
$\textbf{x}=\begin{bmatrix}E_1\\E_2\end{bmatrix}$, $\textbf{A}=\begin{bmatrix}A_1\\A_2\end{bmatrix}$, and
$d\textbf{W}=\begin{bmatrix}dW_1\\dW_2\end{bmatrix}$. The elements of the $\mathbb{B}$ matrix are $\mathbb{B}_{11}=\mathbb{B}_{22}=\sqrt{\beta B/2}N$, $\mathbb{B}_{12}=\mathbb{B}_{21}=0$, and $dW_i=\xi_i(t)dt$. 
Application of Eq.~(\ref{eq:Ito}) to $f(\textbf{x})=E_1^2+E_2^2=P$ gives

\begin{align}
	dP &= \Bigl[\left(\frac {G_N(N-N_t)}{1+\epsilon \mid E\mid^2}-\frac{1}{\tau_p}\right)P+\beta B N^2\Big] dt
	\nonumber
	\\&+ \sqrt{2\beta B}N (E_1dW_1+E_2dW_2). \label{eq:ItoP}
\end{align}
Taking into account that $E_1(t)=\sqrt{P(t)}\cos \phi(t)$, $E_2(t)=\sqrt{P(t)}\sin \phi(t)$, and defining 
\begin{align}
dW_p(t)& = dW_1 (t)\cos \phi (t) + dW_2 (t) \sin \phi (t)\label{eq:ort1}\\
dW_\phi (t)& = -dW_1 (t)\sin \phi (t) + dW_2 (t)\cos \phi (t)
\label{eq:ort2}
\end{align}

we obtain 
\begin{align}   
	d P &=  \Bigl[\left( \frac{G_N(N-N_t)}{1+\epsilon P}  -\frac{1}{\tau_p} \right) P + \beta B N^2 \Bigr]dt\nonumber
	\\&~~~~ + \sqrt{2 \beta BP}N dW_p(t) \label{eq:Pdeducida}
	\end{align}

Eqs.~(\ref{eq:ort1})-(\ref{eq:ort2}) correspond to an orthogonal trasformation in which $dW_p(t)$ and $dW_\phi(t)$ are increments of independent Wiener processes $W_p(t)$ and $W_\phi(t)$  \cite{gardiner1985handbook}. In this way $dW_p(t)=F_p(t)dt$ and $dW_\phi(t)=F_\phi (t)dt$ where
$<F_p(t)>=< F_\phi(t) >=0$, $< F_p(t) F_p(t^\prime)> =< F_\phi(t) F_\phi(t^\prime)> =\delta (t-t^\prime)$, and $< F_p(t) F_\phi(t^\prime)>=0$. Eq.~(\ref{eq:Pdeducida}) is equivalent to  Eq.~(\ref{eq:P}), so it still remains the derivation of Eq.~(\ref{eq:phi}). This is done by applying Eq.~(\ref{eq:Ito}) to 
$f(\textbf{x})=\arctan (E_2/E_1)=\phi$.

\begin{align}
	d\phi &= \frac{1}{P} \left[ -A_1E_2+A_2E_1\right] dt
	\nonumber
	\\&+ \sqrt{\frac{\beta B}{2}}\frac{N}{P} (-E_2dW_1+E_1dW_2). \label{eq:Itophi}
\end{align}

Applying Eqs.~(\ref{eq:A1})-(\ref{eq:A2}), $E_1=\sqrt{P}\cos \phi$, $E_2=\sqrt{P}\sin \phi$, and  Eq.~(\ref{eq:ort2}) we  obtain

\begin{align}
	d\phi &= \frac{\alpha}{2} \left[ G_N(N-N_t) -\frac{1}{\tau_p} \right] dt
	\nonumber
	\\&+ \sqrt{\frac{\beta B}{2P}}N dW_\phi  \label{eq:Itophi1}
\end{align}

that is the same than Eq.~(\ref{eq:phi})

\section{Numerical integration of the field equations}

The Euler-Maruyama algorithm corresponding to Eqs.~(\ref{eq:Efield})-(\ref{eq:Nfield}) can be obtained after splitting Eqs.~(\ref{eq:Efield}) in equations for the real and imaginary part of $E(t)$. The numerical algorithm reads:

\begin{align}
E_1(t+\Delta t) &= E_1(t)+\left(\frac {G_N(N(t)-N_t)}{1+\epsilon  (E_1^2+E_2^2)(t)}-\frac{1}{\tau_p}\right)\frac{E_1(t)}{2}\Delta t
\nonumber
\\& - \alpha\left(G_N(N(t)-N_t)-\frac{1}{\tau_p}\right)\frac{E_2(t)}{2}\Delta t 
\nonumber
\\& + \sqrt{\frac{\beta B\Delta t}{2}}N(t) X_1.
\label{eq:Euler1}
\end{align}

\begin{align}
E_2(t+\Delta t) &= E_2(t)+\left(\frac {G_N(N(t)-N_t)}{1+\epsilon  (E_1^2+E_2^2)(t)}-\frac{1}{\tau_p}\right)\frac{E_2(t)}{2}\Delta t
\nonumber
\\& + \alpha\left(G_N(N(t)-N_t)-\frac{1}{\tau_p}\right)\frac{E_1(t)}{2}\Delta t 
\nonumber
\\& + \sqrt{\frac{\beta B\Delta t}{2}}N(t) X_2.
\label{eq:Euler2}
\end{align}

\begin{align}
	N(t+\Delta t) &= N(t) + \frac{I(t)}{e}\Delta t 
	\nonumber
	\\& -(AN(t)+BN(t)^2+CN(t)^3) \Delta t
	\nonumber
	\\& - \frac{G_N (N(t)-N_t)(E_1^2+E_2^2)(t)}{1+\epsilon(E_1^2+E_2^2)(t) }\Delta t  \label{eq:Euler3}
\end{align}

where $X_1$ and $X_2$ are independent Gaussian numbers with zero mean and standard deviation equal to one. In order to maintain the continuous and unbounded character of the phase within each modulation period, and taking into account that the numerical evaluation of the $\arctan$ function gives values between $-\pi /2$ and $\pi /2$, we have to detect the number of times that the trajectory crosses the vertical axis ($E_1=0$) of the complex plane and the clockwise or counter-clockwise character of these crossings. If the initial condition is such that $E_1(0) >0$ the value of the phase at $t$ is calculated by using:

\begin{equation}
\phi (t)=\arctan \Bigl(\frac{E_2(t)}{E_1(t)}\Bigr)+(n-m)\pi\label{eq:Euler4}
\end{equation}

where $n$ is the total number of counter-clockwise crossings (from quadrant 1, Q1, to quadrant 2, Q2, and from quadrant 3, Q3, to quadrant 4, Q4) observed until time $t$. In the same way $m$ is the total number of clockwise crossings (from Q2 to Q1  and from Q4 to Q3) observed until time $t$. In the opposite case, $E_1(0) < 0$, the value of the phase at $t$ is found from:

\begin{equation}
\phi (t)=\arctan \Bigl(\frac{E_2(t)}{E_1(t)}\Bigr)+(n-m+1)\pi\label{eq:Euler5}
\end{equation}

The previous algorithm has to be modified if we also want to consider the diagonal crossings: from Q1 to Q3, Q3 to Q1, Q2 to Q4, and Q4 to Q2. If we consider that it is equally probable to increase or to decrease the angle in these transitions, we modify the previous algorithm by adding or substracting $\pi$ with probability 1/2 each time one of the previous crossings is observed from $t$ to $t+\Delta t$. We have used this modified algorithm to obtain the results of this work. We note however that the effect of considering these diagonal crossings is much smaller than that obtained when considering only crossings from Q1 to Q2, Q3 to Q4, and viceversa, because the diagonal crossings are just a very small percentage of the total crossings of the vertical axis. For instance, the relative error of the final value of the variance in Fig.~2(b) when using only Eqs.~(\ref{eq:Euler4})-(\ref{eq:Euler5}) with respect to the result found with the modified algorithm is smaller than 1$\%$ for the case $\Delta t=10^{-5}$ ps.

\section{Calculation of the phase variance at the steady state below threshold}

Our departure equations are Eq.~(\ref{eq:Efield}) and a simplified version of Eq.~(\ref{eq:Nfield}) in which linearized recombination of carriers has been considered,

\begin{equation}
\frac{d N}{d t} = \frac{I}{e} - \frac{N}{\tau_n} - \frac{G_N (N-N_t)\mid E\mid^2}{1+\epsilon \mid E\mid^2}  \label{eq:Nfieldap1} 
\end{equation}

where $\tau_n$ is the carrier lifetime. We also consider the situation in which $\alpha =0$. We make these approximations 
in order to obtain simple analytical expressions. They do not affect the main result of the appendix, that is to show in an alternative way the divergence of the optical phase. When the bias current, $I$, is below threshold, $\mid E \mid ^2$ can be neglected and Eqs.~(\ref{eq:Efield})-(\ref{eq:Nfield}) are written as

\begin{equation}
\frac{d E_i}{d t}=a(t)E_i+\sqrt{\frac{\beta B}{2}}N \xi_i(t)\label{eq:Efieldap1}
\end{equation}

\begin{equation}
\frac{d N}{d t} = \frac{I}{e} - \frac{N}{\tau_n} \label{eq:Nfieldap2} 
\end{equation}

where $a(t)=(G_N(N(t)-N_t)-1/\tau_p)/2$, , and $i=1,2$. The solution of  Eq.~(\ref{eq:Nfieldap2}) is 

\begin{equation}
N(t)=\Bigl( N(0)-\frac{\tau_nI}{e}\Bigr) e^{-t/\tau_n} +\frac{\tau_nI}{e}.
\label{eq:Nfieldap3} 
\end{equation}

The solution of Eq.~(\ref{eq:Efieldap1}) is given by 
\cite{valle1992analytical}:

\begin{equation}
E_i(t)=h_i(t)\exp \Bigl( \int_0^t a(s)ds\Bigr)
\label{eq:Efieldap2} 
\end{equation}

where 
\begin{equation}
h_i(t)=E_i(0)+\sqrt{\frac{\beta B}{2}}\int_0^tN(t^\prime)
\exp \Bigl( -\int_0^{t^\prime} a(s)ds\Bigr)\xi_i(t^\prime)dt^\prime.
\label{eq:Efieldap3} 
\end{equation}

$h_i(t)$ are independent Gaussian processes with $< h_i(t) >=E_i(0)$, and variance, $\sigma_i^2= <h_i^2(t)>-<h_i>^2$,
given by:
 
\begin{equation}
\sigma_i^2=\frac{\beta B}{2}\int_0^tN^2(t^\prime)
\exp \Bigl( -2\int_0^{t^\prime} a(s)ds\Bigr)dt^\prime.
\label{eq:Efieldap4} 
\end{equation}

The photon number will be given by:
\begin{equation}
P(t)=E_1^2(t)+E_2^2(t)=\bigl[ h_1^2(t)+h_2^2(t)\bigr]\exp\Bigl( 2\int_0^t a(s)ds\Bigr)
\label{eq:Efieldap5} 
\end{equation}

$P(t)$ is an exponential process because $h_i(t)$ are independent Gaussian processes. The exponential process is determined by just one parameter,  $<P(t)>$, that is calculated by averaging   Eq.~(\ref{eq:Efieldap5}). A simplified expression of $<P(t)>$ can be obtained by considering the evolution at times long enough for $N$ to get a constant value, that is $t>> \tau_n$, $N(t)\approx \tau_nI/e=\bar{N}$, and $a(t)\approx [G_N(\bar{N}-N_t)-1/\tau_p]/2 = \bar{a}$, where $\bar{a}<0$ since $I<I_{th}$. Eq.~(\ref{eq:Efieldap5}) reads

\begin{equation}
P(t)\approx\bigl[ h_1^2(t)+h_2^2(t)\bigr] e^{2\bar{a}t}
\label{eq:Efieldap6} 
\end{equation}

with an average value given by
\begin{equation}
<P(t)>\approx P(0)+\frac{\beta B \bar{N}^2}{2\mid \bar{a}\mid}\bigl( 1-e^{2\bar{a}t}\bigr)
\label{eq:Efieldap7} 
\end{equation}

that becomes independent of time when $t >>\frac{1}{2\mid\bar{a}\mid}$. In the long time regime $P(t)$ becomes an exponential random variable, $P$, with statistical properties independent of time with an average given by $<P>\approx P(0)+\frac{\beta B \bar{N}^2}{2\mid \bar{a}\mid}$.

We now discuss the evolution of the phase by using Eq.~(\ref{eq:Itophi1}), that for $\alpha =0$ reads:

\begin{equation}
d\phi=\sqrt{\frac{\beta B}{2P}}N dW_\phi.
\end{equation}

Integration of this equation gives

\begin{equation}
\phi (t)=\phi (0) +\sqrt{\frac{\beta B}{2}}\int_0^tN(t)\sqrt{\frac{1}{P(t)}} dW_\phi .
\end{equation}

Assuming that $t>>\tau_n$ and $t >>\frac{1}{2\mid\bar{a}\mid}$, we can approximate $N(t)\approx \bar{N}$ and $P(t)\approx P$ so 

\begin{equation}
\phi (t)\approx \sqrt{\frac{\beta B}{2}}\bar{N}\sqrt{\frac{1}{P}}\int_0^t dW_\phi .
\end{equation}

where for simplicity we have assumed that $\phi (0)=0$.  The previous integral is $W_\phi (t)$, a Wiener process with zero mean and variance given by $V[W_\phi (t)]=t$. In this way we have expressed the phase as $\phi (t)=\gamma \sqrt{\frac{1}{P}} W_\phi(t)$ where $\gamma = \sqrt{\frac{\beta B}{2}}\bar{N}$ and where $\sqrt{\frac{1}{P}}$ and $W_\phi (t)$ are statistically independent because $F_P(t)$ and $F_\phi(t)$ are independent. Applying the formula of the variance of the product of two statistically independent random variables we obtain that the variance of the phase is given by:
 
\begin{equation}
\sigma_\phi^2 (t)\approx \gamma^2<\frac{1}{P}>t=\gamma^2t\int_0^\infty \frac{e^{-P/<P>}}{P<P>}dP.
\end{equation}
 
The divergence of the phase variance lies in the fact that this integral has a logarithmic divergence. 



%

\end{document}